\documentclass[aps,prl,twocolumn,groupedaddress,amsmath,amssymb]{revtex4}

\usepackage{graphicx}

\usepackage{color}
\usepackage{bm}
\usepackage{latexsym}

\begin{document}

\title{Interference of sound waves in a moving fluid}

\author{Huanan Li, Andrea Kleeman, Tsampikos Kottos}
\affiliation{Department of Physics, Wesleyan University, Middletown, Connecticut 06459, USA}
\author{Boris Shapiro}
\affiliation{Technion - Israel Institute of Technology, Technion City, Haifa 32000, Israel}

\date{\today}

\begin{abstract}
We investigate sound propagation in a moving fluid confined in a randomly corrugated tube. For weak randomness and small fluid
velocities $v^{(0)}$, the localization length $\xi$ shows extreme sensitivity to the variation of $v^{(0)}$. In the opposite limit of large 
fluid velocities, $\xi$ acquires a constant value which is independent of the frequency of the incident sound wave, the degree of 
randomness and $v^{(0)}$ itself. Finally, we find that the standard deviation $\sigma_{\ln T}$ of the logarithm of transmittance $\ln(T)$ 
is a universal function of the ensemble average $\langle \ln T\rangle $, which is not affected by the fluid velocity.
\end{abstract}


\maketitle
{\it Introduction -}
Wave transport in random media is the focus of many theoretical and experimental studies during the last sixty years \cite{50years}. One 
of the fascinating phenomena that has been predicted and subsequently observed in such media is the halt of wave propagation:
it was found that due to multiple scattering and the consequent destructive interference between scattered waves, the total transmittance 
decays exponentially with the size of the system. This phenomenon, known as Anderson localization, has been originally predicted 
in  the realm of condensed matter physics \cite{A58,LR69}. More recently it has been studied and observed in optics \cite{WBLR97,
LAPSMCS08,PPKSBNTL04,SBFS07}, microwaves \cite{CSG00,BZKKS09}, acoustics \cite{acoustic}, as well as for matter waves in cold atoms
systems \cite{SL10,S12}.

In the present paper we investigate Anderson localization in a new setting, namely, sound propagation in a {\it moving fluid}. For simplicity
we assume that the fluid is confined to a one-channel waveguide with random corrugation. We consider subsonic flows where the fluid
is inviscid and turbulent effects are irrelevant. We find that wave interference that lead to Anderson localization of sound are strongly
affected by the velocity of the flow $v^{(0)}$.  In a broad range of parameter (strength of disorder, wave frequency, size and cross-section
of the scatterers) the localization length $\xi$ is extremely sensitive to $v^{(0)}$, as long as $v^{(0)}$ is not too large. As $v^{(0)}$ increases,
$\xi$ saturates at a universal value, independent of the wave frequency. We also find that the variance $\sigma_{\ln T}^2$ of the logarithm 
of transmittance $\ln(T)$ is a universal function of its average value $\langle\ln T\rangle$, independent of the  flow velocity $v^{(0)}$.

{\it Mathematical modeling - } We consider sound propagation in a tube with a moving fluid. Moreover we will focus our analysis on one-
dimensional propagation which imposes the constraint that the wavelength of the sound is much larger than the typical width of 
the tube. The tube consists of two parts: the left $x<-L/2$ and the right $x>L/2$ (semi-infinite) domains have a constant cross section 
$A_0$ and constitute the ``leads" from where the sound is emitted and detected respectively, while in the domain $-L/2<x<L/2$ the tube 
cross-section $A(x)$ is non-uniform (corrugated domain). We will assume that the fluid flows from left to right with a constant velocity 
$v^{(0)}$ at the leads.

The one-dimensional equations for mass and momentum conservation read \cite{Landau}:
\begin{eqnarray}
\label{cons}
A(x){\partial \rho(x,t)\over \partial t} +{\partial \over \partial x}\left[\rho(x,t) v(x,t) A(x)\right]=0; &\\
{\partial v(x,t)\over\partial t} +v(x,t) {\partial v(x,t)\over \partial x}+{1 \over \rho(x,t)}{\partial p(x,t)\over \partial x} = 0&\nonumber
\end{eqnarray}
where $\rho(x,t)$ denotes the fluid density, $v(x,t)$ is its velocity, and $p(x,t)$ is the pressure. In what follows we assume that the corrugated
region consists of uniform segments, with cross-section $A_n$ for the $n$-th segment. The segments are separated by sharp transition 
regions in which the cross-section rapidly changes from $A_n$ to $A_{n+1}$. The precise profile of various quantities in the transition
regions are complicated and we eliminate those regions by imposing boundary conditions which are obtained by integrating Eqs. (\ref{cons})
across the transition regions between adjacent segments. This results in the continuity of the quantities: $\rho v A$ and ${1\over 2} v^2+w$, 
where $w$ is the enthalpy per unit mass. While integrating the second equation in Eq. (\ref{cons}) we have used the relation
$(dw)_s = ({1\over \rho}) {dp}$ which is valid for isentropic flow (the entropy $s$ per unit mass is constant).

All the quantities in Eq. (\ref{cons}), for each segment $n$, have a stationary (time-independent) part $\{p^{(0)},\rho^{(0)}, v^{(0)}\}$ upon which
small oscillatory terms $\{p'(x,t),\rho'(x,t), v'(x,t)\}$ are superimposed, due to the sound wave. Linearizing Eqs. (\ref{cons}) with respect to the
oscillatory terms yields, in the $n-$th segment
\begin{equation}
\label{eqlin}
{Dp_n'\over Dt} +c_0^2\rho^{(0)} {\partial v_n'\over \partial x}=0;\quad {Dv_n'\over Dt} +{1\over \rho^{(0)}} {\partial p_n'\over \partial x}=0
\end{equation}
where ${D\over Dt} \equiv \left({\partial \over \partial t} + v_n^{(0)} {\partial \over \partial x}\right)$ is the so-called convective derivative and
$c_0\equiv \sqrt{(\partial p/\partial \rho)_s}$ the adiabatic speed of sound in the reference frame of the fluid. In deriving Eqs. (\ref{eqlin}) we
have used the relation $p_n'=c_0^2\rho_n'$. Moreover, to somewhat simplify the treatment, we have assumed a nearly incompressible fluid
so that $c_0$ is the same in each segment. Eliminating $v_n'$ from Eqs. (\ref{eqlin}) leads to the equation for the pressure wave
in a moving fluid \cite{RH14}
\begin{equation}
\label{eqp}
{D^2p_n'\over Dt^2} = c_0^2 {\partial^2p_n'\over \partial x^2},
\end{equation}
whose solution can be written in terms of two counter-propagating waves:
\begin{eqnarray}
\label{solution1}
p_n'(x,t)&=& P_n^f e^{ik_n^fx-i\omega t} + P_n^b e^{-ik_n^bx-i\omega t},
\end{eqnarray}
where $\omega=k_n^f(c_0+v_n^{(0)})=k_n^b(c_0-v_n^{(0)})$ and $k_n^f (k_n^b)$ indicate the wave vector of the waves propagating in (opposite to)
the direction of the flow.
Correspondingly, the velocity variation $v_n'(x,t)$ as determined from Eq.~(\ref{eqlin}) is
\begin{eqnarray}
\label{solution2}
v_n'(x,t)&=&{1\over \rho^{(0)} c_0} \left(P_n^fe^{ik_n^fx-i\omega t} - P_n^be^{-ik_n^bx-i\omega t}\right).
\end{eqnarray}

At the boundary between two adjacent segments the values of $v'$ and $p'$ should be matched by the boundary conditions. The latter are
obtained by linearization of the two aforementioned continuity conditions, namely for $\rho vA$ and for ${1\over 2} v^2+w$. This results
in continuity, accross the boundary, of the stationary quantities $v^{(0)} A $, and ${1\over 2} (v^{(0)})^2+w_0$, and of the oscillatory quantities
$\rho^{(0)} (v' A)+(v^{(0)} A) \rho'$, $v^{(0)}v'+{1\over \rho^{(0)}} p'$. The last two conditions can be conveniently re-written as
\begin{eqnarray}
\label{bound2}
{1\over \rho^{(0)}} p_n' + v_n^{(0)} v_n' &=& {1\over \rho^{(0)}} p_{n+1}' + v_{n+1}^{(0)} v_{n+1}', \\
\left[1-\left({v_{n}^{(0)}\over c_0}\right)^2\right] A_n v_n' &=& \left[1-\left({v_{n+1}^{(0)}\over c_0}\right)^2\right] A_{n+1} v_{n+1}' \nonumber
\end{eqnarray}
At the left and right leads, where the cross-section $A_0$ is constant, the pressure and velocity variations are given
by Eqs.~(\ref{solution1}) and (\ref{solution2}) with the sub-indexes $n=L(R)$ for the left (right) lead, and $v_L^{(0)}=v_R^{(0)}=v^{(0)}$.
In the corrugated domain $-L/2<x<L/2$, substitution of Eqs.~(\ref{solution1}) and (\ref{solution2}) into Eqs.~(\ref{bound2}) allows us to cast the 
boundary relations into a transfer matrix form which connects the forward and backward propagating sound wave amplitudes 
between the two subsequent domains, $n$ and $n+1$.

Finally, the pressure at $x=L/2$ is related to that at $x=-L/2$ via the total transfer matrix ${\mathcal M}$:
\begin{eqnarray}  \label{transfer0}
\left(\begin{array}{c}
P_{R}^{f} e^{ik^f L/2}  \\
P_{R}^{b} e^{-ik^b L/2} \end{array}
\right) & = & \mathcal{M} \left(\begin{array}{c}
P_{L}^{f} e^{ik^f (-L/2)}   \\
P_{L}^{b} e^{-ik^b (-L/2)} \end{array}
\right).
\end{eqnarray}

The transmission and reflection amplitudes for left and right incident waves can be expressed in terms of the transfer matrix elements as
$t_L={\det{\mathcal M}\over {\mathcal M}_{22}}; r_L=-{{\mathcal M}_{21}\over {\mathcal M}_{22}}$ ($t_R={1\over {\mathcal M}_{22}}; r_R=
{{\mathcal M}_{12}\over {\mathcal M}_{22}}$). These relations have been obtained from Eqs.~(\ref{transfer0}) by imposing the appropriate 
scattering conditions $P_R^b=0$ ($P_L^f=0$) associated to left (right) incident waves respectively. Furthermore, straightforward calculations 
show that $|\det{\mathcal M}|=1$,  which means that $|t_L|=|t_R|=|t|$.

Transmittance $T$ and reflectance $R$ are defined as ratios of the corresponding energy fluxes (currents), ${\cal I}$ \cite{current}. The latter 
are proportional to the square of the wave amplitudes, for instance, the current for the forward propagating wave in the left lead is ${\cal I}^f_L 
\sim |P^f_L|^2$.  However, the proportionality coefficients for the forward and backward propagating waves are different. Since transmittance 
involves a pair of waves (incident and transmitted) propagating in the same direction, one simply has $T_L=|t_L|^2$ for the transmittance from 
left to right and, similarly, $T_R=|t_R|^2$ for transmittance in the opposite direction (the two are the same due to the previously mentioned 
equality $|t_L|=|t_R|$). On the other hand, reflectance involves a pair of waves (incident and reflected) propagating in the opposite directions 
which leads to the relations $R_L=\gamma \cdot |r_L|^2$ and  $R_R=(1/\gamma)\cdot |r_R|^2$. The coefficient $\gamma=\left({c_0-v^{(0)}
\over c_0+v^{(0)}}\right)^2$ is due to the fact that the reflected waves travel with a different speed than the incident waves. Armed with the 
above knowledge we are now ready to investigate the effects of interference due to scattering from defects.

\begin{figure}[h]
\includegraphics[width=1\columnwidth,keepaspectratio,clip]{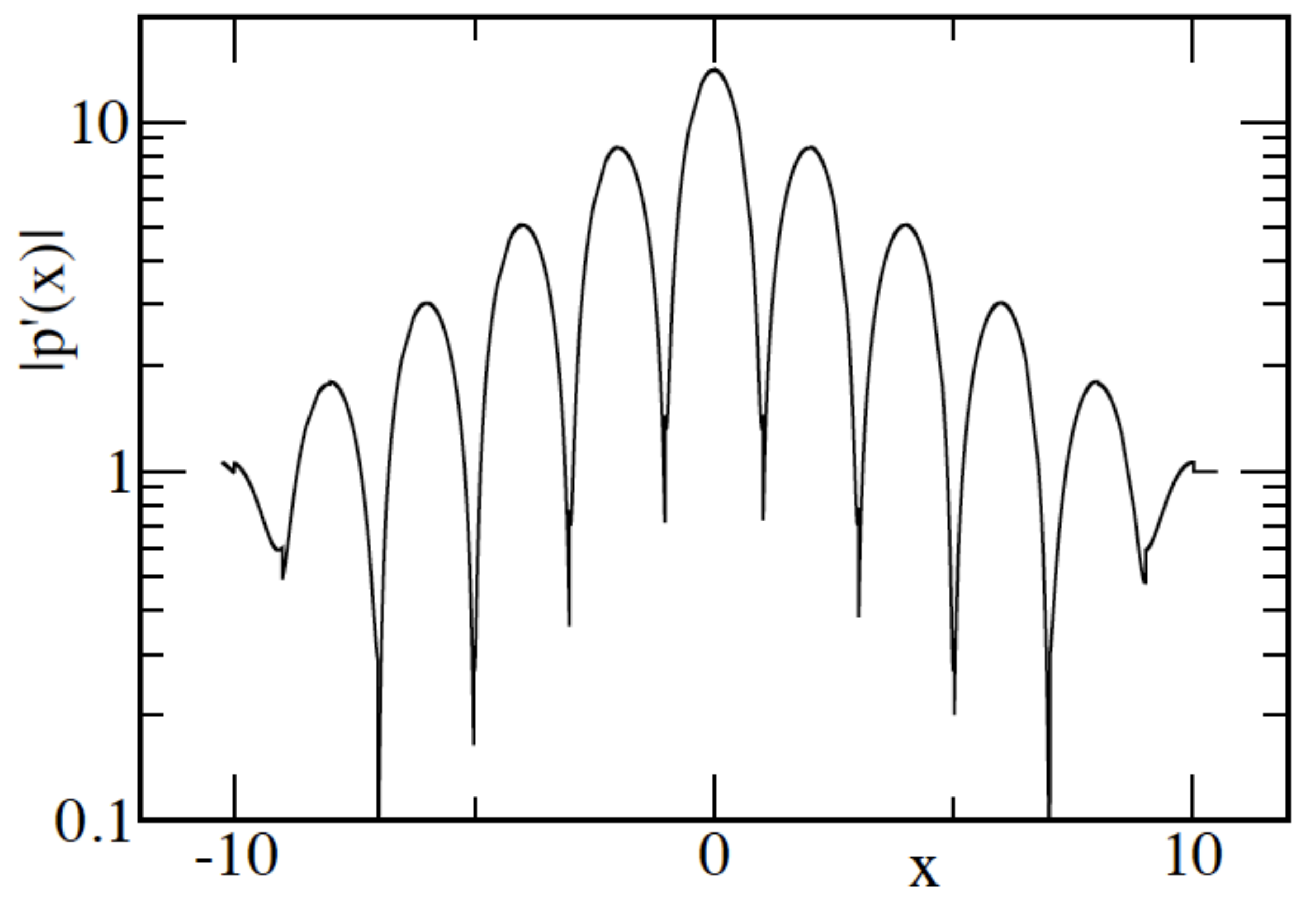}
\caption{Pressure profile $p'(x)$ along a periodically modulated tube with corrugated areas characterized by $L_1=L_2, A_1=0.6 A_2$ 
and a defect $L_d=0.05 L_2,A_d=A_2$ in the middle of the tube (we consider $L_2=1, A_2=1$). A localized mode is created due to 
interference effects between the periodic modulation and the defect. The cross section of the leads is $A_0=0.6A_2$ and the velocity of the 
fluid is $v^{(0)}=0.1 c_0.$
}
\label{fig1}
\end{figure}

{\it One Defect --} It is instructive to start our analysis with the simple example of one defect embedded in an otherwise uniform
infinite tube of cross section $A_0$. The cross section of the defect is $A_d$ and it occupies the interval $-L_d/2<x<L_d/2$. First
we note that, regardless of the ratio $A_d/A_0$, a single defect cannot support a bound state. This is because Eq. (\ref{eqp}) in the
leads, away from the defect,  cannot have exponentially decaying solutions for real $\omega$, i.e. the dispersion relation requires
real $k$ for real $\omega$. The situation here is different from that in quantum mechanics, where an attractive potential can produce
a bound state with a negative energy (imaginary $k$). However when the defect is placed in a periodically modulated tube consisting 
of segments with lengths $L_1$, $L_2$ and cross-sections  $A_1,  A_2$ then a localized defect state can be formed (see 
Fig. \ref{fig1}). This is a result of interference between the periodic corrugation and the defect.

Next we study the transmittance properties of a single defect. The  boundary conditions Eqs.~(\ref{bound2}) at $x=
\pm L_d/2$ allow us to evaluate the transfer matrix ${\cal M}_d$ and consequently the transmittance $T_d$. We
obtain:
\begin{equation}
\label{Tdefect}
T_d={1\over 1+2 \eta \left[1-\cos\left(\phi_d\right)\right]};\,\,\eta=({1-\alpha^2\over 4 \alpha})^2
\end{equation}
where $\alpha={A_0\over A_d}$, $\phi_d={2 L_d\omega/c_0\over 1-\beta_d^2}$ and $\beta_d=v_d^{(0)}/c_0$. From Eq. (\ref{Tdefect}) 
we see that the resonance modes of the defect (corresponding to $T_d=1$)
are achieved when $\phi_d=2\pi m$ (where $m=1, 2,\cdots$). The resonance
condition can be re-written in a more transparent way as $(k_d^f+k_d^b) L_d= 2\pi m$ which resembles the standard
resonance condition, albeit now the two counter-propagating waves have different wave-vectors. Finally, it
is important to realize that $\phi_d$ depends not only on the incident frequency $\omega$ but also on the velocity
of the flow $\beta_d$ which can lead to strong changes in transmittance.

{\it Disorder Case --} We proceed with the analysis of the transport properties of the sound in a disordered tube. We therefore destroy
the periodicity of the corrugated tube consisting of two different cross sections $A_1$ and $A_2$ by uniformly randomizing their associated
lengths $L_1, L_2$ in such a way that $L_1\in [L_1^{(0)}-\delta_1, L_1^{(0)}+\delta_1]$ and $L_2\in [L_2^{(0)}-\delta_2, L_2^{(0)}+\delta_2]$.
In our simulations below we will use $A_0=A_1=1, A_2=1.2$. The transmittance $T$ has
been evaluated using the transfer matrix formalism Eq. (\ref{transfer0}). 

From the Anderson theory of localization we expect that for long enough tubes (and/or strong disorder) the transmittance will decay
exponentially with the size of the disordered sample $L$ (associated with $N\approx L/(L_1^{(0)}+L_2^{(0)})$ number of scattering units). 
This exponential decay is best described by the rescaled localization length 
${\tilde \xi}\equiv \xi/(L_1^{(0)}+L_2^{(0)})$ which is defined as:
\begin{equation}
\label{loclength}
{\tilde \xi}^{-1}\equiv -\lim_{L\rightarrow \infty} {L_1^{(0)}+L_2^{(0)}\over L}\langle\ln (T)\rangle.
\end{equation}
where $\langle \cdots\rangle$ indicates an averaging over disorder realizations. All results 
have been averaged over more than 200 different disorder realizations.

\begin{figure}[h]
\includegraphics[width=1\columnwidth,keepaspectratio,clip]{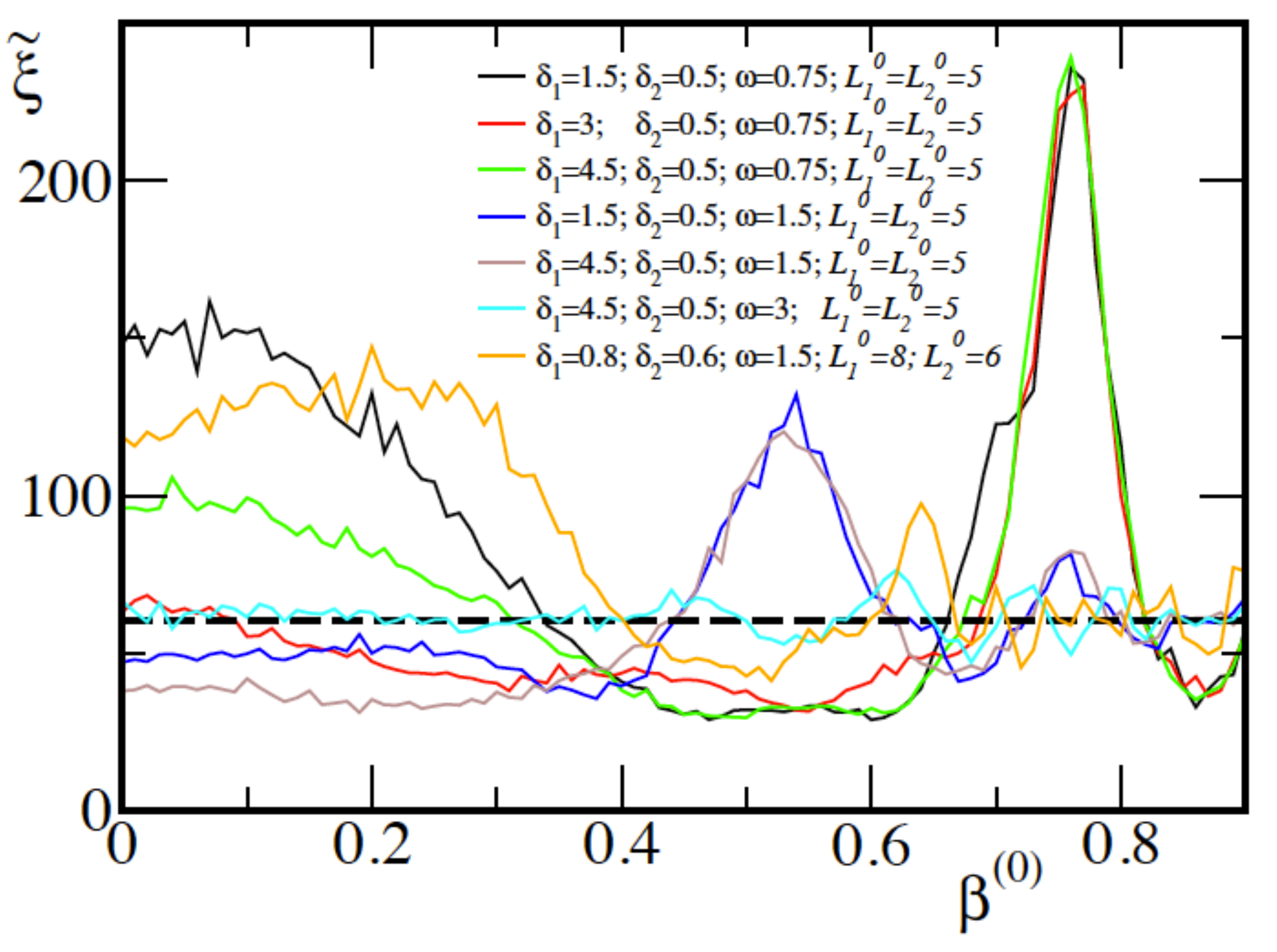}
\caption{Dimensionless localization length ${\tilde \xi}=\xi/(L_1^{(0)}+L_2^{(0)})$ versus the rescaled lead fluid velocity  
$\beta^{(0)}=v^{(0)}/c_0$. Various lines correspond to different disordered strengths $\delta$, mean corrugation width $L_1^{(0)}, 
L_2^{(0)}$ and different frequencies $\omega$. The horizontal black dashed line is the result of RPA Eq. (\ref{RPA}).
}
\label{fig2}
\end{figure}

The dependence of ${\tilde \xi}$ on the scaled fluid velocity $\beta^{(0)}=v^{(0)}/c_0$ for various disordered strengths 
$\delta_1,\delta_2$, mean lengths $L_1^{(0)}$, $L_2^{(0)}$ and different frequencies is reported in the main part of Fig. \ref{fig2}. Our 
results indicate that for small velocities $v^{(0)}\leq 0.4 c_0$, the localization length is sensitive to the variations of $v^{(0)}$ while for
larger values of $v^{(0)}$ it originally oscillates and finally saturates to a universal value which is independent of $\omega, \delta_1,
\delta_2$, and $v^{(0)}$ itself.

An understanding of the universal value of the rescaled localization length is achieved by employing a random phase approximation (RPA).  
We consider a tube consisting of $(N-1)$ scattering units to which we add one more scattering unit. The transmission amplitude $t_L^{(L)}$
of the combined system is
\begin{eqnarray}
\label{ttot}
t_L^{(N)}&=&t_L^{(N-1)} \left(1+r_L^{(1)}e^{ik_1^bL_1}r_R^{(N-1)}e^{ik_1^fL_1}+\cdots\right)e^{ik_1^fL_1}t_L^{(1)}\nonumber\\
                 &=&{t_L^{(N-1)} e^{ik_1^fL_1}t_L^{(1)}\over 1- r_L^{(1)}e^{i(k_1^f+k_1^b)L_1}r_R^{(N-1)}}
\end{eqnarray}
where the subscript $(N-1)$ designate the transmission and reflection amplitudes for the $(N-1)$ chain, while superscrips $(1)$ (and $L_1$)
refer to the added scattering unit.
Writing $r_L^{(1)}=\sqrt{\frac{R_L^{(1)}}{\gamma}} e^{i\phi_{r_L^{(1)}}}$ and $r_R^{(N-1)}=\sqrt{\gamma R_R^{(N-1)}} 
e^{i\phi_{r_R^{(N-1)}}}$, yields the following expression for the logarithm of the total transmittance $T_L^{(N)}=|t_L^{(N)}|^2$
\begin{eqnarray}
\label{Tlog}
\ln T_L^{(N)}&=&\ln T_L^{(1)} +\ln T_L^{(N-1)} \\
&-&\ln\left[1+ R_L^{(1)} R_R^{(N-1)}-2\sqrt{R_L^{(1)}R_R^{(N-1)}} \cos\Phi\right]\nonumber
\end{eqnarray}
where $\Phi=(k_1^f+k_1^b)L_1+\phi_{r_L^{(1)}}+\phi_{r_R^{(N-1)}}$ and $T_L^{(1)}$ is given by Eq. (\ref{Tdefect}) with the substitution 
$(A_d,L_d, \phi_d)\rightarrow (A_2,L_2,\phi_2)$. Furthermore, if one assumes that randomness in $L_1, L_2$ is such that the associated 
phases $\phi_{1,2}$ are completely randomized (i.e. uniformly distributed between $-\pi$ and $\pi$), then averaging over these  
phases yields the rescaled (inverse) localization length
\begin{equation}
\label{RPA}
{\tilde \xi}^{-1} = \ln\left[{1\over 2} (1+2\eta + \sqrt{1+4\eta})\right].
\end{equation}
Thus, in the RPA the (rescaled) localization length ${\tilde \xi}$ (see Eq. (\ref{loclength})) depends only on the ratio $A_1/A_2$ (recall the definition
of $\eta$, Eq. (\ref{Tdefect})). The RPA Eq. (\ref{RPA}) is also plotted in Fig. \ref{fig2} and matches our numerical results in the large 
$v^{(0)}$ domain.

\begin{figure}[h]
\includegraphics[width=1\columnwidth,keepaspectratio,clip]{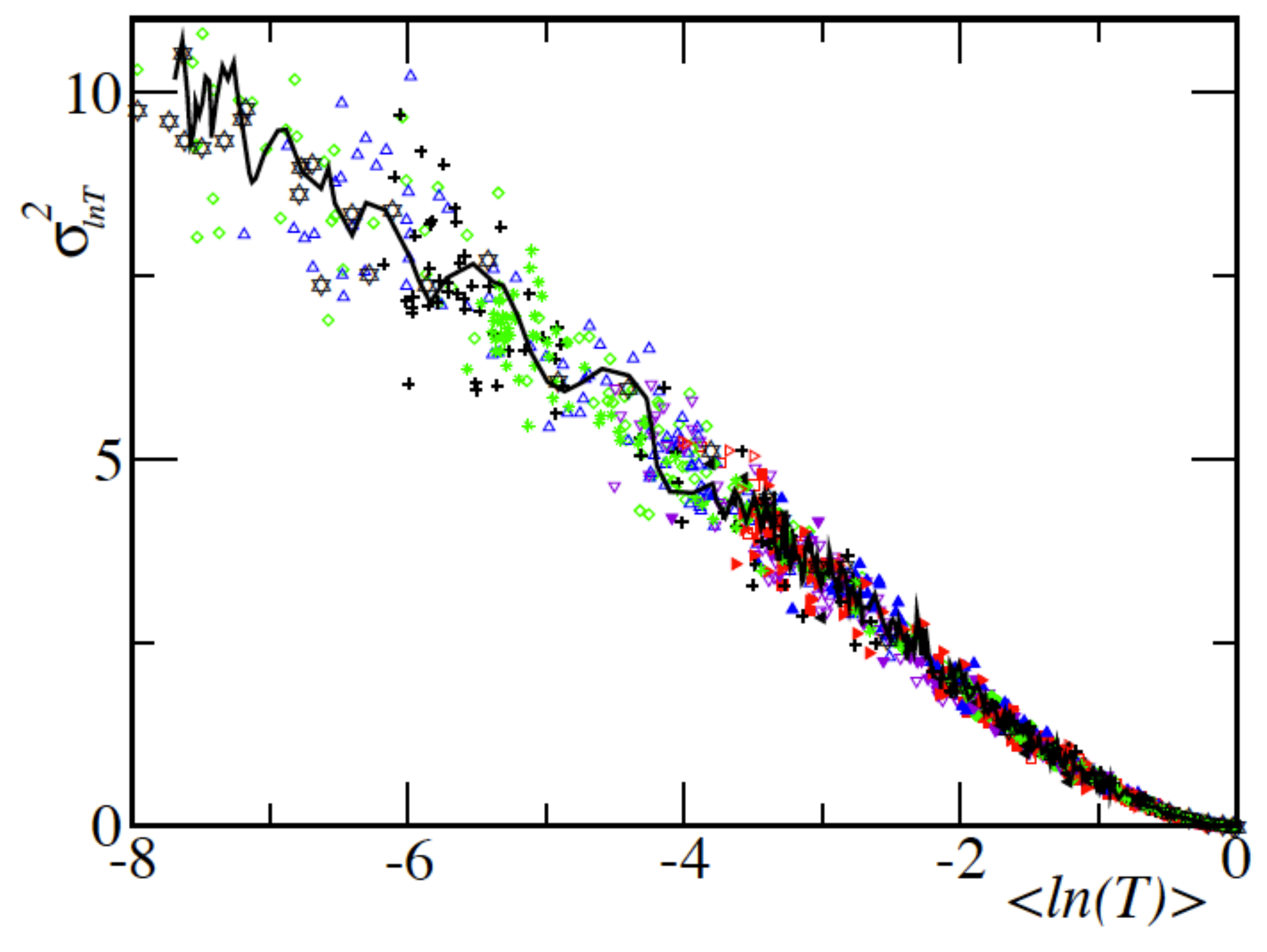}
\caption{Variance of logarithm of transmittance $\sigma_{\ln T}^2$ versus its mean value $\langle \ln T\rangle$ for various 
disordered strengths $\delta_1,\delta_2$, mean widths $L_1^{(0)}, L_2^{(0)}$, frequencies, and velocities $v^{(0)}$ in the interval 
$[0,0.8]$. All data collapse to one universal curve characterizing also the $v^{(0)}=0$ case (black solid line).
}
\label{fig3}
\end{figure}


The sensitivity of $\xi$ to $v^{(0)}$ for small fluid velocities (see Fig. \ref{fig2}) can be understood by considering the averaging over 
the random lengths $L_1,L_2$ more carefully. We first assume that $\phi_2$ is not random (all lengths $L_2=L_2^{(0)}$ are identical) 
while the randomness in $L_1$ is such that $\delta \phi_1=2(\omega/c) \delta_1/(1-\beta_1^2)>\pi$. In this case, the propagation 
phases at the $A_1$ sections are completely randomized so that one can expect $\langle \ln T^{(N)}\rangle_1 \approx N \ln T_2$ where 
$T_2$ is given by Eq. (\ref{Tdefect}) with the obvious substitution $(L_d,A_d,\phi_d)\rightarrow (L_2, A_2,\phi_2)$ ($\langle\cdots 
\rangle_1$ indicates an average over $\phi_1$-phases). In the case of $\eta\ll 1$, the rescaled localization length reads
\begin{equation}
\label{rloc}
{\tilde \xi}^{-1} \approx 2\eta \left[1-\cos(\phi_2)\right]
\end{equation}
which indicates that when $\beta_2$ is changing, ${\tilde \xi}^{-1}$ exhibits (aperiodic) oscillations. The $m-$th oscillations is 
completed when $\beta_2$ takes the values $\beta_{2}^{(m)}=\left(1+{L_2^{(0)}\omega\over c_0\pi m}\right)^{-1/2}$.

Next we introduce randomness also into the phase $\phi_2$, i.e. we assume that $\phi_2=\phi_2^{(0)}+\Delta \phi_2$ where $\phi_2^{(0)}=
2(\omega/c_0)L_2^{(0)}/(1-\beta_2^2)$ and $\Delta \phi_2=2(\omega/c_0)\Delta L_2/(1-\beta_2^2)$. 

As long as $\Delta\phi_2\ll1$,  the disorder in $L_2$ has only a minor effect and ${\tilde \xi}$ is approximately given by Eq. (\ref{rloc}), 
with $\phi_2$ replaced by its average value $\phi_2^{(0)}$. There is though a small correction, related to the variance of $L_2$, 
so that
\begin{equation}
\label{decay}
{\tilde \xi}^{-1}=2\eta\left[1-\cos(\phi_2^{(0)}) + {2\over 3}\cos(\phi_2^{(0)}) 
{\left({\omega\over c_0} \delta_2^2\right)^2 \over(1-\beta_2^2)^2}\right].
\end{equation}
We should note that the above relation is not applicable when $\beta_2\rightarrow 1$ since in this case $\Delta\phi_2$ is not small. Moreover
Eq. (\ref{decay}) indicates that when $\phi_2^{(0)}\ll 1$ then ${\tilde \xi}^{-1}\approx 4\eta (\omega L_2^{(0)}/c_0)^2/(1-\beta_2^2)^2$ i.e.
the localization length decreases as $\beta_2=\alpha \beta^{(0)}$ increases. In the opposite case of $\phi_2^{(0)}\gg 1$ a decrease or growth of  
${\tilde \xi}$ as $\beta_2$ increases from zero (and while $\beta_2\ll 1$) depends on the sign of $\sin(2\omega L_2^{(0)}/c_0)$ (see Fig. 
\ref{fig2}). This condition can be obtained from the expansion of $\cos(\phi_2^{(0)})$ for small $\beta_2$ and after neglecting the small 
contributions from the last term in Eq. (\ref{decay}). Finally when $\Delta \phi_2\gg1$ we recover the RPA also for the segments $L_2$. In this case $\langle 
\cos(\phi_2)\rangle_2=0$ in Eq. (\ref{rloc}) and we obtain the results of Eq. (\ref{RPA}) for $\eta\ll 1$.


We have also analyzed the fluctuations of the transmission for various disordered strengths $\delta_1,\delta_2$, system sizes $L$ and
fluid velocities $v^{(0)}$. Our numerical results for $\sigma^2_{\ln T}\equiv \langle (\ln T)^2\rangle -\langle \ln T\rangle^2$
versus $\langle \ln T\rangle$ are reported in Fig \ref{fig3} for various values of $v^{(0)}\in [0, 0.8]$. We find that, despite the sensitivity
of $\xi$ to the fluid velocity (for small values of $v^{(0)}$), the variance $\sigma^2_{\ln T}$ is a universal function of $\langle \ln T\rangle
=L/\xi$, independent of the velocity of the fluid. In fact, the analysis indicates that this universal function is identical to the one associated
with the $v^{(0)}=0$ case (black bold line in Fig. \ref{fig3}). Our results indicate strong fluctuations in the localized regime due to interference 
among multiple scattered sound waves.

{\it Conclusions --}In conclusion, we study sound waves propagating ``on top'' of a  
stationary fluid flow, in the presence of disorder. It turns out that  
the stationary flow can have a significant effect on the interference  
pattern of the waves, as compared to the case when the fluid is at  
rest. This happens because the speed of sound (in the laboratory  
frame), and therefore the phases accumulated between scattering  
events,  depend on the propagation direction of the wave, i.e.  along  
the flow or opposite to it.  We find that in a broad range of  
parameters the localization length $\xi$ is very sensitive to the  flow  
velocity $v^0$. When $v^0$ increases, $\xi$ can grow, diminish or oscillate-  
depending on the precise value of other parameters. However for large  
$v^0$ (comparable to the speed of sound) phases become completely  
randomized and $\xi$ saturates at some universal value, in agreement with  
the ``random phase approximation''.

\begin{acknowledgments}

\end{acknowledgments}



\begin{thebibliography}{99}
\bibitem{current} The current along the $x$-direction is defined as ${\cal I}=(\rho_0 v' +\rho' v_0)({c_0^2 \rho'\over \rho_0} + v_0 v')$ \cite{RH14}.
\end{thebibliography}
\end{document}